\documentclass[%
 aip,
 reprint,%
]{revtex4-1}

\usepackage{graphicx}

\begin{document}


\title{Strong-Field Terahertz Control of Plasmon Induced Opacity in Photoexcited Metamaterial}



\author{Ali Mousavian}
\affiliation{Department of Physics, Oregon State University, Corvallis, OR 97331-6507, USA}
\author{Zachary J. Thompson}
\affiliation{Department of Physics, Oregon State University, Corvallis, OR 97331-6507, USA}
\author{Byounghwak Lee}
\affiliation{Department of Physics, Oregon State University, Corvallis, OR 97331-6507, USA}
\affiliation{Department of Physics and Chemistry, Korea Military Academy, Seoul 01805, South Korea}
\author{Alden N. Bradley}
\author{Milo X. Sprague}
\author{Yun-Shik lee}
\email{leeys@physics.oregonstate.edu}
\affiliation{Department of Physics, Oregon State University, Corvallis, OR 97331-6507, USA}


\date{\today}

\begin{abstract}
A terahertz metamaterial consisting of radiative slot antennas and subradiant complementary split-ring resonators exhibits plasmon induced opacity in a narrow spectral range due to the destructive interference between the bright and dark modes of the coupled oscillators. Femtosecond optical excitations instantly quench the mode coupling and plasmon oscillations, injecting photocarriers into the metamaterial. The plasmon resonances in the coupled metamaterial are restored by intense terahertz pulses in a subpicoseond time scale. The strong terahertz fields induce intervalley scattering and interband tunneling of the photocarries, and achieve significant reduction of the photocarrier mobility. The ultrafast dynamics of the nonlinear THz interactions reveals intricate interplay between photocarriers and plasmon oscillations. The high-field THz control of the plasmon oscillations implies potential applications to ultrahigh-speed plasmonics.
\end{abstract}


\maketitle 



High performance terahertz (THz) devices are in high demand as THz applications in a variety of fields have emerged in recent years.~\cite{Lee:09,Yu:19,Ma:19} A major obstacle to developing THz devices of high functionality is that naturally occurring materials generally have flat spectral responses in the THz region. THz metamaterials, however, can be artificially engineered to have resonances at specific frequencies.~\cite{Chen:06,ChoiM:11,Ju:11} The various patterns of metamaterial ensure versatile functions to manipulate THz radiation. THz metamaterials have been used as absorbers, modulators, filters, switches, lenses, etc.~\cite{Grant:11,Chen:09,Nemec:09,Chen:07,Neu:10,Yang:14}

A plasmonic metamaterial consisting of radiative elements coupled with subradiant elements exhibits plasmon induced transparency (PIT), where a sharp transmission resonance emerges within a broad absorption band due to destructive interference between the bright and dark modes.~\cite{Zhang:08,Kekatpure:10,Artar:11,zhu:13,Liu:12} The extreme dispersion of PIT in a narrow spectral region signifies high sensitivity of the metamaterial to external perturbations.~\cite{Li:11,Gu:12}

\begin{figure}[ht]
\center{\scalebox{0.46}{\includegraphics{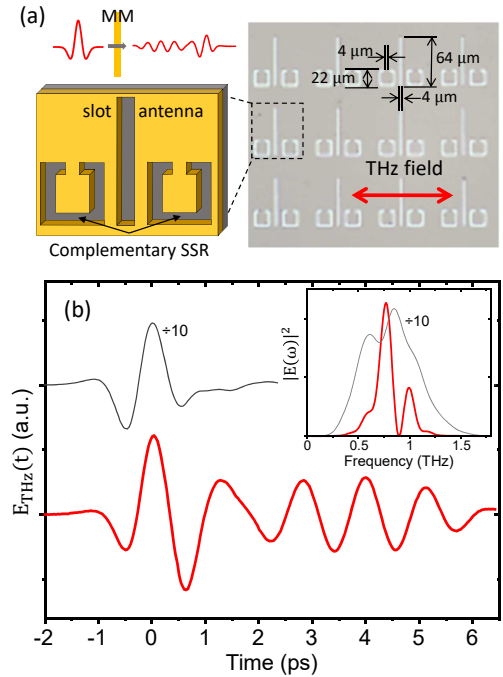}}}
\caption{a) Unit cell and optical microscope image of the THz metamaterial consisting of radiative slot antennas (length, 64~$\mu$m; width, 4~$\mu$m) and subradiant CSRRs (length, 22~$\mu$m; gap, 4~$\mu$m) (b) THz waveform transmitted through the metamaterial (red) The incident single-cycle THz waveform is in gray (the field amplitude is reduced by a factor of 10). Spectra of the waveforms are in inset.} \label{fig1}
\end{figure}

In this work we study nonlinear THz effects in a metamaterial of a complementary PIT structure. Figure~\ref{fig1}a shows a sketch of the unit cell consisting of a slot antenna and two complementary split-ring resonators (CSRR) and a microscope image of the antenna array. Incident THz fields perpendicular to the slot-antenna axis radiatively couple to the antenna, while the CSRRs are subradiant for the THz polarization. The Babinet’s principle indicates that the complementary structure has the same resonant frequency with the opaque structure, while their transmission spectra are inverted from each other, i.e., $T_c(\omega)=1-T_o(\omega).$~\cite{Horak:19} The complementary structure has an advantage that its transmitted light contains radiation only from the resonators having no background of the incident light.


The THz metamaterial was fabricated by etching out the plasmonic elements in a 500-nm-thick Al film deposited on an intrinsic GaAs substrate. The THz waveform transmitted through the metamaterial (red line in Fig.~\ref{fig1}b) shows plasmon oscillations when a single-cycle THz pulse (gray line in Fig.~\ref{fig1}b) excites the coupled resonators. Fitting the data with a numerical simulation of a coupled harmonic oscillator model produces the resonant frequencies ($\nu_b$=0.85~THz and $\nu_d$=0.90~THz) and the damping rates ($\gamma_b/2\pi$=0.15~THz and $\gamma_d/2\pi$=0.03 THz) of the bright and dark modes. The dark mode decays five times slower than the bright mode. The mode coupling is strong enough that the coupling coefficient, $\nu_c$=0.42 THz, is comparable to the mode frequencies. The spectrum of the transmitted light (red line in the inset of Fig.~\ref{fig1}b) presents a sharp opaque dip embeded in a broad peak. The narrow resonance of the dark elements (CSRRs) emerges in the radiation spectrum due to the coupling between the bright mode of slot-antennas and the dark mode of CSRRs. The plasmon induced opacity (PIO) is the complementary phenomenon of PIT.

We performed time-resolved high-field THz spectroscopy of the THz metamaterial upon femtosecond optical excitation. We employed tilted-pulse-front optical rectification in a Mg:LiNbO$_3$ prism to generate intense single-cycle THz pulses. The light source was a 1-kHz regenerative amplifier (wavelength, 800~nm; pulse energy, 1~mJ; pulse duration, 100~fs). THz pulses were focused to near diffraction limit onto the sample using off-axis parabolic mirrors (beam waist, 0.4~mm). The field amplitude of the broadband THz pulses (central frequency, 0.9~THz; bandwidth, 0.9~THz) reached 1~MV/cm at an optical pulse energy of 1~mJ. We traced the THz waveforms with electro-optic sampling in a 1-mm ZnTe crystal. The incident THz fields, perpendicular to the slot-antenna axis, radiatively couple to the slot antennas, while the CSRRs become dark elements in this polarization configuration. A small portion of the regenerative amplifier output was used for optical excitation of the metamaterial.


\begin{figure}[ht]
\center{\scalebox{0.75}{\includegraphics{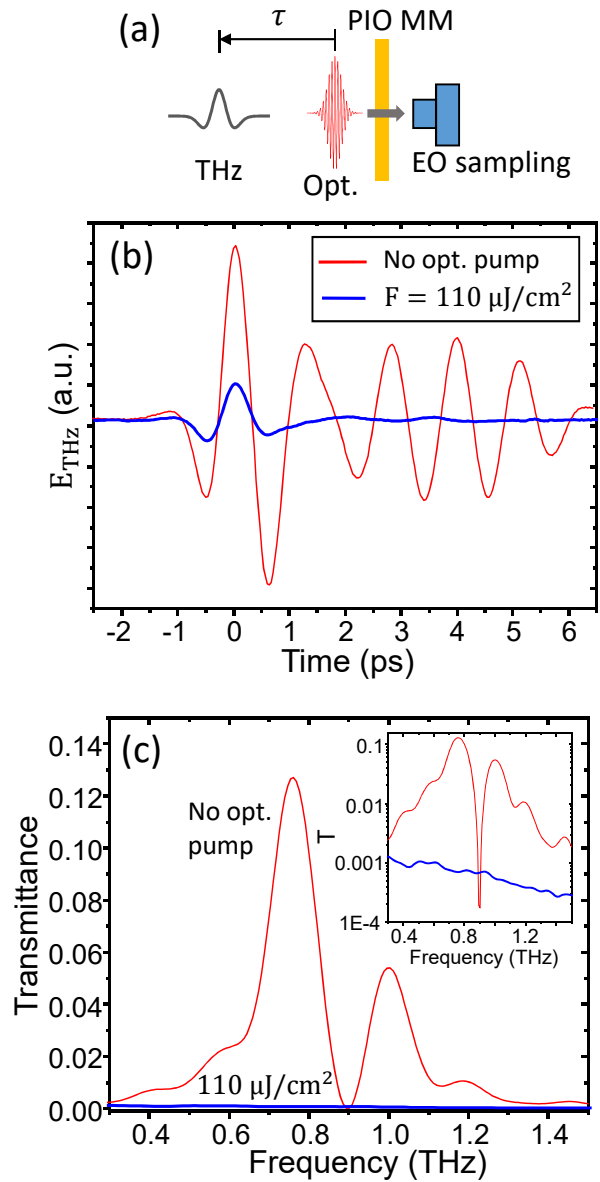}}}
\caption{(a) Schematic diagram of the optical-pump THz-probe experiment. (b) THz waveforms transmitted through the metamaterial with (blue) and without (red) optical excitation (pump fluence, $F$=110~$\mu$J/cm$^2$; time delay between the optical and THz pulses, $\tau$=-150 ps). The peak field amplitude of the incident THz pulse $E_p$ is 85~kV/cm. (c) Transmittance spectra of the waveforms. Spectra in log-scale are shown in the inset.} \label{fig2}
\end{figure}

We examine the PIO metamaterial with a weak THz probe when the sample is optically excited by femtosecond laser pulses (Fig.~\ref{fig2}). An ultrafast optical excitation turns off the plasmon resonances of the metamaterial, injecting photocarriers in the insulating region of the composite media. The THz pulses transmitted through the optically excited sample (blue line in Fig.~\ref{fig2}b) are very small and show no plasmon oscillations. The waveform of no optical excitation is shown in red for comparison. The transmittance spectrum (blue line in Fig.~\ref{fig2}c) presents the collapse of the plasmon resonances. The log-scale spectra in the inset of Fig.~\ref{fig2}c more vividly illustrate the optical quenching of the plasmon resonances. The photoexcitation flattens the spectrum almost entirely, yet it raises transmittance in the narrow frequency range where the opacity dip of no photoexcitation is extremely deep. The optical control of PIO demonstrates novel applications to THz sensing and modulation of high fidelity.~\cite{Li:11,Gu:12}


\begin{figure}[ht]
\center{\scalebox{0.8}{\includegraphics{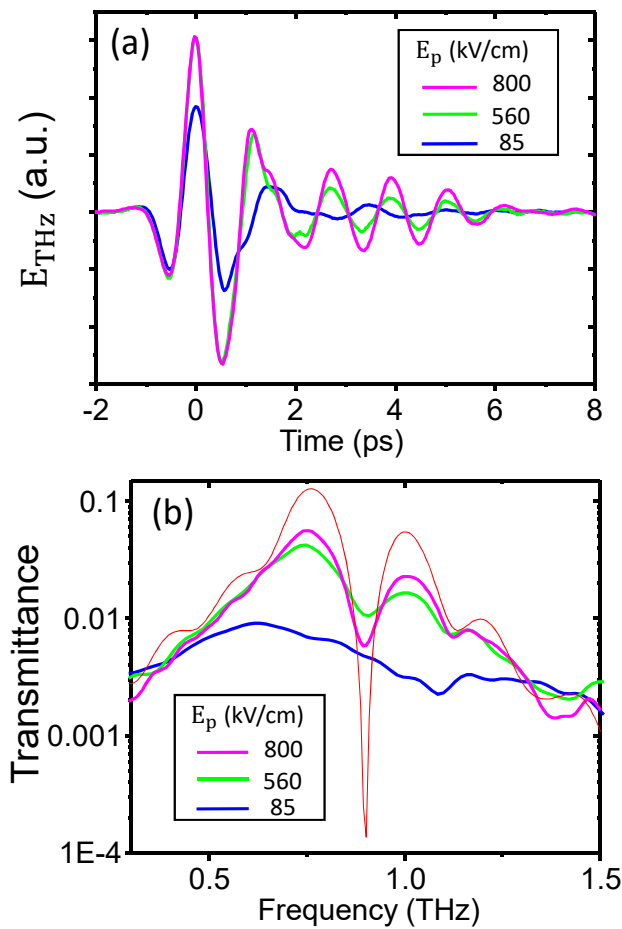}}}
\caption{Optical-pump THz-control experiment: (a) Transmitted THz waveforms for the peak field strength of the incident THz pulse, $E_p$=800 (violet), 560 (green) and 85 (blue)~kV/cm (optical pump fluence, $F$=12~$\mu$J/cm$^2$; time delay, $\tau$=-150 ps) (b) Transmittance spectra of the waveforms in log-scale. The spectrum of no optical excitation is in red.} \label{fig3}
\end{figure}

We employ strong THz pulses to control the plasmon resonances in the photoexcited metamaterial altering the mobility of photocarriers. High-field THz pulses have been used to manipulate photocarriers in semiconductors,\cite{Sharma:12} which has potential applications to ultrahigh-speed optoelectronics.~\cite{Mousavian:18,Mousavian:18a} Figure~\ref{fig3}a shows that intense THz fields restore plasmon oscillations in the optically excited metamaterial, where the oscillation amplitudes increase with the THz field strength. The opacity dip in the transmittance spectra grows as the THz intensity increases, while the spectrum of the weak THz pulse is flat and small (Fig.~\ref{fig3}b). The instant revival of the plasmon oscillations is accounted for by the conductivity decrease in GaAs due to the reduction of electron mobility. The dominant mechanisms of the mobility reduction in photoexcited GaAs are field-driven intervalley scattering and interband tunneling.~\cite{Mousavian:18,Lange:14}


\begin{figure}[ht]
\center{\scalebox{0.6}{\includegraphics{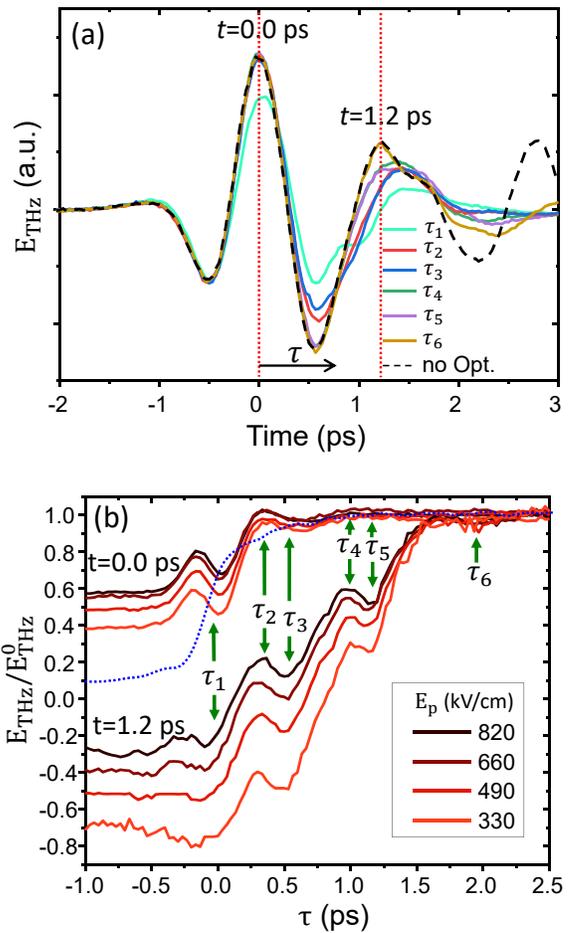}}}
\caption{(a) Transmitted THz waveforms at time delays between optical and THz pulses, $\tau_{1-6}$=-0.07, 0.37, 0.53, 1.00, 1.17, and 1.93~ps for $F$=61~$\mu$J/cm$^2$ and $E_p$=820~kV/cm. (b) Normalized transmitted THz field $E_{THz}(t)/E_{THz}^0(t)$ ($E_{THz}^0$: field with no optical pump) at $t$=0.0 and 1.2~ps versus time delay $\tau$ for the peak field strength of the incident THz pulse, $E_p$=820, 660, 490 and 330~kV/cm. The dotted blue line is $E_{THz}/E_{THz}^0$ through a bare GaAs substrate at $t$=0.0~ps.} \label{fig4}
\end{figure}

The high-field dynamics of photocarriers are swayed by the plasmon oscillations in the metamaterial. We observe the ultrafast dynamics of the interplay between the plasmon oscillations and photocarriers monitoring the THz transmission as a function of the time delay $\tau$ between the optical and THz pulses near $\tau=0$. Figure~4a shows the transmitted THz waveforms at $\tau$=-0.07, 0.37, 0.53, 1.00, 1.17, and 1.93~ps for the peak field strength of the incident THz pulse, $E_p=820$~kV/cm, when the optical pump fluence is 61~$\mu$J/cm$^2$. The main peak at $t=0.0$~ps is severely attenuated at $\tau_1$=-0.07 ps, while it changes little at the positive time delays, $\tau_{2-6}$. Figure~\ref{fig4}b shows that the amplitude of the main peak at $t=0$ widely fluctuates between $\tau$=-0.5 and 0.5~ps, exhibiting an oscillatory behavior in the time window. On the contrary, the THz field transmitted through a bare GaAs substrate (dotted blue line in Fig.~\ref{fig4}b) shows no oscillation, indicating that the THz interactions with photocarriers are influenced by the plasmon resonances. In Fig.~\ref{fig4}a, the second peak at $t$=1.2~ps undergoes more complex modulations consisting not only of attenuation, but also of phase shift and broadening. Figure~\ref{fig4}b presents the oscillatory modulations of the second peak as a function of $\tau$. The swift and large-scale fluctuations of the transmitted THz fields, which are absent in a bare GaAs substrate, imply that the photoconductivity rapidly undulates as the plasmon fields sway the photocarriers between different sidebands.

In summary, the THz metamaterial consisting of radiative slot antennas and subradiant CSRRs exhibits an extremely deep opacity band at the resonant frequency of the dark element. Femtosecond laser excitations instantly quench the plasmon oscillations in the coupled THz metamaterials, while strong THz pulses restore the sharp plasmonic resonance on a subpicosecond time scale. The effects of high-field THz pulses on the photoexcited plasmonic metamaterial demonstrate the ultrafast dynamics of photocarriers interacting with plasmon oscillations. The experimental scheme presents potential applications to ultrahigh-speed control of metamaterial and plasmonic devices.

Acknowledgements: This work was supported by the National Science Foundation (DMR-1905634).

\bibliographystyle{prsty}

\end{document}